\documentclass[fleqn,usenatbib,letters]{mnras}

\usepackage{newtxtext,newtxmath}

\usepackage[T1]{fontenc}

\DeclareRobustCommand{\VAN}[3]{#2}
\let\VANthebibliography\thebibliography
\def\thebibliography{\DeclareRobustCommand{\VAN}[3]{##3}\VANthebibliography}

\usepackage{graphicx}	
\usepackage{amsmath}	

\DeclareUnicodeCharacter{2212}{-}

\title[Compact Object DM SNe~Ia]{Type Ia supernova  constraints on compact object dark matter}

\author[S. Dhawan et al.]{
S. Dhawan,$^{1}$\thanks{E-mail: sd919@cam.ac.uk}
E. M{\"o}rstell$^{2}$
\\
$^{1}$Institute of Astronomy and Kavli Institute for Cosmology, University of Cambridge, Madingley Road, Cambridge CB3 0HA, UK\\
$^{2}$Oskar Klein Centre, Department of Physics, Stockholm University, Albanova University Center, SE-106 91 Stockholm, Sweden\\
}

\date{Accepted XXX. Received YYY; in original form ZZZ}

\pubyear{2015}

\begin{document}
\label{firstpage}
\pagerange{\pageref{firstpage}--\pageref{lastpage}}
\maketitle

\begin{abstract}
The nature of dark matter (DM) is an open question in cosmology, despite its abundance in the universe. While elementary particles have been posited to explain DM, compact astrophysical objects such as black holes formed in the early universe offer a theoretically appealing alternate route. Here, we constrain the fraction of DM that can be made up of primordial black holes (PBHs) with masses $M \gtrsim 0.01 M_\odot$, using the Type Ia supernova Hubble diagram. Utilizing the Dyer-Roeder distance relation, where the homogeneous matter fraction is parameterized with $\eta$, we find a maximum fractional amount of DM in compact objects ($f_p$) of 0.50 at 95\% confidence level (C.L.), in the flat $\Lambda$CDM model and 0.49 when marginalising over a constant dark energy equation of state. These limits do not change when marginalising over cosmic curvature, demonstrating the robustness to the cosmological model. When allowing for the prior on $\eta$ to include $\eta > 1$, we derive $f_p < 0.32$ at 95$\%$ C.L., showing that the prior assumption of $\eta \leq 1$ gives a conservative upper limit on $f_p$. When including Cepheid calibrated supernovae, the 95\% C.L. constraints improve to $f_p < 0.25$. We find that the estimate for the Hubble constant in our inference is consistent with the homogeneous case, showing that inhomogeneities in the form of compact dark matter cannot account for the observed Hubble tension. In conclusion, we strongly exclude the possibility that PBHs with stellar masses and above form a dominant fraction of the dark matter. 
\end{abstract}



\vspace{-0.6cm}
\section{Introduction}
The nature of dark matter (DM), which constitutes close to 25\% of the energy content of the universe and sources the formation of large scale structure (LSS), is poorly understood. A leading scenario is a new elementary particle beyond the standard model of particle physics, e.g. axions, or WIMPs produced in the early universe  having suppressed interaction with standard model particles  \citep[e.g.][]{Bertone2005}. Cosmological observations are insensitive to the microscopical nature of DM as long as it behaves as a non-relativistic fluid, i.e. cold dark matter (CDM) on large scales. Contrary to the microscopic DM scenario, massive astrophysical compact objects, e.g. primordial black holes (PBHs) produced in the early universe \citep{Zeldovich1967,Carr1974},  have been proposed as DM candidates. PBHs behave as non-relativistic matter on large scales, making them viable as CDM candidates. Depending on their properties, such as their mass, they can be probed by small-scale effects, e.g. the imprint on the magnitude-redshift relation of luminosity distance indicators \citep[e.g.][]{Metcalf1999}.

Type Ia supernovae (SNe~Ia) are excellent cosmological distance indicators, instrumental in the discovery of accelerated expansion of the Universe and local measurements of the expansion rate; the Hubble constant, \citep[$H_0$;][]{Riess2022}. SNe~Ia are also important tracers of the large scale structure of the universe. They have been used to measure the growth of structure, $\sigma_8$ \citep{Quartin2014,Stahl2021}, with the expectation that future surveys will constrain it at few percent precision. 
While explanations of late-time acceleration usually assume the cosmological principle, i.e. homogeneity and isotropy at large ($\sim$ 100 Mpc),  inhomogeneities at smaller scales impact the SN~Ia magnitude-redshift relation \footnote{In this study, we only use relative distances for SNe~Ia, hence we use the terms magnitude-redshift relation and Hubble diagram interchangeably}.
Hence, the SN~Ia magnitude-redshift relation is also sensitive to the fraction of DM in compact objects, e.g. PBHs \citep[e.g.][]{Moertsell:2001ah,ZS2018,Dhawan2018c}. 

An approximation to account for light traveling through emptier rather than denser regions of the universe was proposed in \citet{Zeldovich1964} where a correction to the luminosity-redshift relation in homogeneous models, known as the Dyer-Roeder  (DR) relation, was formulated \citep{Dyer1972,Dyer1973}. In this \emph{ansatz}, the total matter density, $\rho_{\rm m}$ contributes to the expansion rate of the universe, whereas only a fraction of the density, $\eta \cdot \rho_m$, contributes to the gravitational focusing of the light along the line of sight 
to faraway point sources, e.g. Type Ia supernovae (SNe~Ia). Here, $\eta$ is referred to as the the homogeneity parameter. 
This approximate distance-redshift relation allows us to measure the inhomogeneity of the universe in compact form through, $(1-\eta) / \alpha$, where $\alpha$ is the correction term from the Weyl focusing as well as possible Ricci focusing from non-compact dark matter inhomogeneities, such as smooth galactic halos. This correction term is computed from modelling the matter distribution, however, as shown in \cite{Mortsell2002}, it is fairly insensitive to the details of the modelling, indicating $\alpha \sim 0.6$. There is renewed interest in compact objects as DM candidates since the detection of gravitational waves (GW) from binary black hole (BBH) detected by the LIGO-Virgo Collaboration \citep[LVC;][]{Abbott2016,Abbott2022}. Additionally, the mass range of these detections coincides with the range with the fewest robust abundance constraints.

In this paper, we use the DR distance relation to constrain the amount of DM that can feasibly be made up of compact objects. 
The inference of $\eta$ is model independent, and as discussed above, the transformation to $f_p$ is also similar for different model parameter assumptions. This simplicity of the DR approximation allows us to constrain the amount of dark matter in compact objects while simultaneously constraining properties of dark energy, e.g., its density and equation of state \citep[e.g.,][]{Dhawan2018c}, as well as exploring other, more complicated cosmologies, e.g. including spatial curvature ($\Omega_{\rm K}$). Complementary approaches for constraining $f_p$ have been developed in the literature, using the magnification probability distribution function \citep[PDF; e.g.][]{ZS2018}. This approach impact the Hubble residuals by shifting the peak towards fainter residuals since most sightlines would be demagnified compared to the mean and a long tail towards high magnification for the sightlines passing close to the compact objects  \citep[e.g.][]{SH99,Metcalf1999,ZS2018}. It had previously not been feasible to derive meaningful constraints on $f_p$ using the DR distance approach. However, here we present the first stringent constraints on compact object dark matter while robustly marginalising over dark energy properties. These constraints on $f_p$ are sensitive to compact objects where the Einstein radius is larger than the size of the SN~Ia photosphere, i.e. $M \gtrsim 0.01\,{\rm M_\odot}$ \citep[e.g., see][]{Carr_2022}, the mass range observed by LVC. Our method is presented in section~\ref{sec:method} and results in section~\ref{sec:results}. We discuss our results in context of the literature and conclude in section~\ref{sec:discuss}. 
\vspace{-0.6cm}
\section{Data and Methodology}
\label{sec:method}
Here we describe the formalism for inferring the cosmological parameters along with $\eta$, the homogeneous matter fraction. 
The general differential equation for the distance between two light rays of the boundary of a small light cone propagating far away from all clumps of matter in an inhomogenous universe is developed in \citet{1964SvA.....8...13Z, 1965SvA.....8..854D,1966SvA.....9..671D,1967ApJ...150..737G}. This relation for the angular diameter distance given in \cite{Dyer1973} is

\begin{equation}
Q D_{\rm A}^{\prime\prime} + (\frac{2Q}{1+z} + \frac{Q^{\prime}}{2}) D_{\rm A}^{\prime} + \frac{3}{2} \eta\, \Omega_\mathrm{M}\, (1+z)\, D_{\rm A} = 0 ,
\label{eq:dr_dist}
\end{equation}
where $\eta$ is the fraction of the matter density homogeneously distributed and $Q(z) = E^2(z)$, hence, $Q(z)$ depends on the cosmological model \citep[e.g., see][]{2013PhRvD..87f3527B,Helbig2015a}. Below we describe the dataset and formalism used for parameter inference in this study.

\subsection{Type Ia supernovae}
For our analysis we use the most recent SN~Ia magnitude-redshift relation from the latest SN~Ia compilation, i.e., Pantheon+ \citep{Brout2022}. 

Theoretically, the distance modulus predicted by the homogeneous and isotropic, flat Friedman-Robertson-Walker (FRW) universe is given by

\begin{equation}
\mu(z; \theta) = 5\, \mathrm{log_{10}} \left( \frac{d_{\rm L}}{10\, \mathrm{Mpc}} \right) + 25
\label{eq:mu_sne}
\end{equation}
where $z$ is the redshift, $\theta$ are the cosmological parameters and $d_{\rm L}$ is given by 
\begin{table}
\caption{CMB compressed likelihood parameters, errors and covariance matrix for each model fit in this analysis}
    \centering
    \begin{tabular}{|c|c|c|c|c|c|}
    \hline
    $\Lambda$CDM & Parameter Value & $R$ & $l_{\rm A}$ & $\Omega_{\rm b} h^2$ \\
      $R$   & 1.7502 $\pm$ 0.0046    & 1  &  0.46  &  -0.66 \\   
     $l_{\rm A}$ & $301.471^{+0.089}_{-0.090}$  & 0.46  &  1 & −0.33  \\
     $\Omega_{\rm b} h^2$ & 0.02236 $\pm$ 0.00015 & -0.66 & -0.33 & 1 \\
     \hline \\
     $w$CDM & Parameter Value & $R$ & $l_{\rm A}$ & $\Omega_{\rm b} h^2$ \\
      $R$  & 1.7493 $^{+0.0046} _{-0.0047}$    & 1  &  0.47  &  -0.66 \\
     $l_{\rm A}$ & 301.462$^{+0.089}_{-0.090}$  & 0.47  &  1 & −0.34  \\
     $\Omega_{\rm b} h^2$ & 0.02236 $\pm$ 0.00015 & -0.66 & -0.34 & 1 \\
     \hline
     o$\Lambda$CDM & Parameter Value & $R$ & $l_{\rm A}$ & $\Omega_{\rm b} h^2$ \\
      $R$   & 1.7429 $\pm$  0.0051  &   1  &  0.54  &  -0.75 \\ 
     $l_{\rm A}$ & 301.409 $\pm 0.091$  & 0.54  &  1 & −0.42  \\
     $\Omega_{\rm b} h^2$ & 0.0226 $\pm$ 0.00017 & -0.75 & -0.42 & 1 \\
     \hline
    \end{tabular}
    \label{tab:cmb_compress}
\end{table}

\begin{equation}
D_{\rm L} = D_{\rm A} \cdot (1 + z)^2
\label{eq:lum_dist}
\end{equation}
using the Etherington distance duality relation \citep{Etherington1933} and $D_{\rm A}$ from equation~\ref{eq:dr_dist}, wherein $Q = E^2$,  

\begin{equation}
E^2(z)   = \Omega_\mathrm{M} (1+z)^3 + \Omega_{\mathrm{DE}}(z) + \Omega_\mathrm{K}(1+z)^2, 
\label{eq:norm_hubbleparameter}
\end{equation}
where 

\begin{equation}
\mathrm{\Omega_{\mathrm{DE}}(z)} = \Omega_{\mathrm{DE}}\, \mathrm{exp} \left[ 3 \int_0^z \frac{1+w(x)}{1 + x} dx \right],
\label{eq:ode_z}
\end{equation}
$w(z)$ is the dark energy EoS. In this study, we test models with $w$, being constant with redshift, as a free parameter.

Observationally, the distance modulus is calculated from the SN~Ia peak apparent magnitude ($m_B$), light curve width ($x_1$) and colour ($c$)

\begin{equation}
\mu_{\rm obs} = m_B - (M_B - \alpha x_1 + \beta c) + \Delta_\mathrm{M} + \Delta_\mathrm{B},
\label{eq:obs_distmod}
\end{equation}
where $M_B$ is the absolute magnitude of the SN~Ia. Here, $\alpha$, $\beta$, are the slopes of the width-luminosity and colour-luminosity relation. $\Delta_\mathrm{M}$ and $\Delta_\mathrm{B}$ are the host galaxy mass step and distance bias corrections respectively \citep[see, ][for details ]{Brout2022}.

The $\chi^2$ is given by 
\begin{equation}
\chi_{\mathrm{SN}}^2 = \Delta^T C_{\mathrm{SN}}^{-1} \Delta, 
\end{equation}
where $\Delta = \mu - \mu_{\rm obs}$ and $C_{\mathrm{SN}}$ is the complete covariance matrix described in \cite{Brout2022}.

\subsection{Complementary datasets}
\subsubsection{Cosmic Microwave Background}
Precise constraints on the expansion history can be derived by combining the SN~Ia magnitude-redshift relation with complementary cosmological probes. For the complementary probes, we assume $\eta = 1$ since the   average angular diameter distance at a fixed redshift over the large regions probed by the CMB and the baryon acoustic oscillations converge to the distance in a homogeneous universe \citep[e.g.,][]{Kibble2005, Metcalf2007}. 
For the geometric constraints from the CMB we use the compressed likelihood from the \emph{Planck} satellite, \cite[see][, and Table~\ref{tab:cmb_compress}, for details]{Chen2019}. The CMB shift, $R$, position of the first acoustic peak in the power spectrum, $l_{\rm A}$ and the baryon density at present day, $\Omega_{\rm b} h^2$ comprise the data vector. The expression for the CMB shift and the position of the first acoustic peak are given by

\begin{equation}
R = \sqrt{\Omega_\mathrm{M} H_0^2} d_{\rm A}(z_*)/c,
\label{eq:cmb_shift}
\end{equation}
and
\begin{equation}
l_{\rm A} = \pi \frac{d_{\rm A}(z_*)}{r_s(z_*)}
\label{eq:cmb_la}
\end{equation}
where $r_s(z)$ is the sound horizon at redshift, $z$, given by 
\begin{equation}
r_s(z) = \frac{c}{\sqrt{3}} \int^{\frac{1}{1+z}}_0 \frac{da}{a^2 H(a) \sqrt{1 + a\frac{3\Omega_\mathrm{b}}{4\Omega_\gamma}}}, 
\label{eq:sound_horizon}
\end{equation}
where $\Omega_\gamma = 2.469\cdot 10^{-5} h^{-2}$ and $h= H_0/100$.

For the parameter values, uncertainties and
the elements of the covariance matrix $C_{ij} = \sigma_i \sigma_j D_{ij}$. Differences in covariance matrix for the various models  are small.
\vspace{-0.3cm}
\subsubsection{Baryon Acoustic Oscillation}
The detection of the characteristic scale of the BAO in the correlation function of different matter distribution tracers provides a powerful standard ruler to probe the angular-diameter-distance versus redshift relation. BAO analyses use a few different distance measurements depending on the dataset, namely, $D_{\rm A}$, $D_{\rm C}$, and $D_{\rm V}$ which are related as 
\begin{equation}
D_{\rm A} = \frac{D_{\rm C}}{1+z}
\label{eq:da}
\end{equation}

\begin{figure}
    \centering
        \includegraphics[width=.48\textwidth]{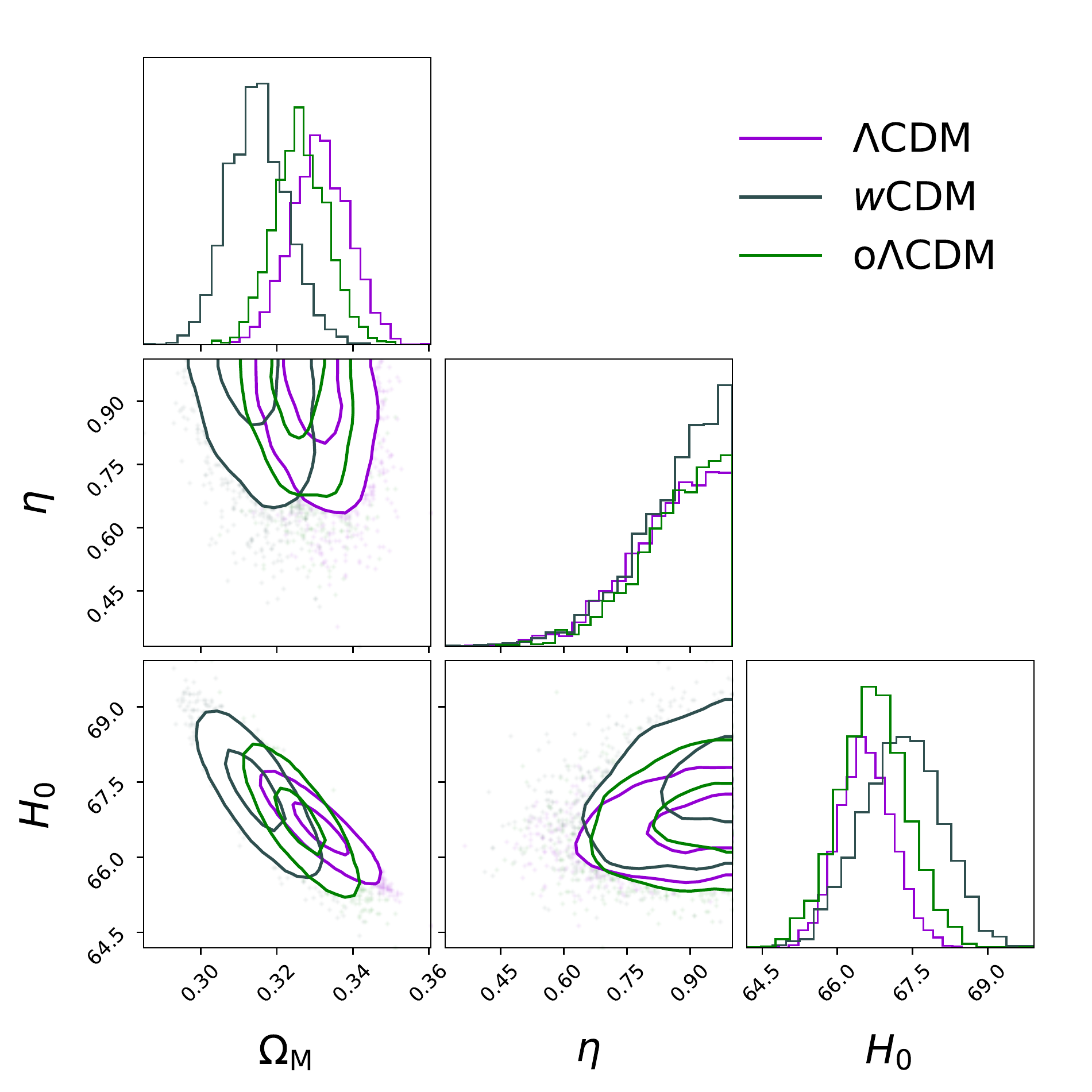}
    \caption{Constraints on $\Omega_{\rm M}$, $\eta$ and $H_0$ for all the models tested here, namely, F$\Lambda$CDM, $w$CDM and o$\Lambda$CDM. We find only a weak degeneracy between $\Omega_{\rm M}$ and $\eta$ and the 95$\%$ C.L. for $\eta$ is consistent across the models. }
    \label{fig:eta_allmodels}
\end{figure}
\begin{equation}
D_{\rm V}(z) = \left[(1+z)^2 D_{\rm A}(z)^2 \frac{cz}{H(z)} \right] ^{1/3}, 
\label{eq:dvz}
\end{equation}
where $H(z)$ is the Hubble parameter at redshift, $z$. Cumulatively, we refer to these measurements as $d_x$.

For our analyses, we use measurements from 6dFGS \citep{Beutler2011}, SDSS MGS \citep{Ross2015}, BOSS DR12 \citep{bossdr12}, BOSS DR14 \citep{Bautista2018}, eBOSS QSO \citep{Zhao2019}, spanning a range of redshifts $z ]\in [0.106, 1.944]$, considering a BAO likelihood of the form 
\begin{equation}
\chi^2_{\mathrm{BAO}} = (d_x - d_x^{\mathrm{BAO}})^T C_{\mathrm{BAO}}^{-1} (d_x - d_x^{\mathrm{BAO}}). 
\label{eq:chi_bao}
\end{equation} 
We refer to \citet{Hogas2021} for more details on the data sets.  

\vspace{-0.4cm}
\section{Results}
\label{sec:results}
\begin{table*}
    \centering
    \caption{Parameter constraints for the models tested in this work.}
    \begin{tabular}{|c|c|c|c|c|c|c|c|}
    \hline
    Model & $\Omega_{\rm M}$ & $f_p$ &  $\Omega_{\rm b} h^2$ & $H_0$ & $w$ & $\Omega_{\rm K}$\\
    \hline
F$\Lambda$CDM & 	0.3313 $\pm$  0.0076 & 	0.50 & 	0.0222 $\pm$  0.0002 & 66.565 $\pm$  0.539 & $\ldots$ & $\ldots$ \\
$w$CDM & 	0.3147 $\pm$  0.0078 & 	0.49 & 	0.0226 $\pm$  0.0002 & 67.298 $\pm$  0.798 & -0.951 $\pm$  0.038 & $\ldots$ \\
o$\Lambda$CDM & 	0.3261 $\pm$  0.0071 & 	0.47 & 	0.0226 $\pm$  0.0002 & 66.730 $\pm$  0.689 & $\ldots$ & -0.0022 $\pm$  0.0021 \\
F$\Lambda$CDM ($\eta$ = U[0, 5]) & 0.3081 $\pm$ 0.0071 & 0.32 & 0.0225 $\pm$ 0.0002 & 68.202 $\pm$ 0.566 & $\ldots$ & $\ldots$ \\
F$\Lambda$CDM (With Calibrators) & 0.2952 $\pm$ 0.0062 & 0.25 & 0.0225 $\pm$ 0.0002 & 69.179 $\pm$ 0.498 & $\ldots$ & $\ldots$ \\
\hline
    \end{tabular}
    \label{tab:constraints}
\end{table*}
We fit the standard $\Lambda$CDM model to the combined dataset from the SNe~Ia, BAO and CMB to constrain $\Omega_{\rm M}$ and $\eta$. With a uniform prior on $\eta$ of [0,1] we find that the data constrains $\eta$ to be close to one, with a 95$\%$ C.L. lower limit on $\eta$ of 0.74. Using the conversion between $\eta$ and $f_p$ from \citep{Mortsell2002}, i.e. $f_p = \frac{(1 - \eta - 0.02)}{0.653}$ we get $f_p < 0.50$ at the 95$\%$ C.L. The denominator in the previous expression includes the correction for the Weyl focusing (and possible Ricci focusing from smooth matter structures). We note that this is a conservative upper limit, as $f_p$ is the amount of \emph{all} matter in compact objects, not only DM. The constraints on the amount of baryonic matter in stars is $\sim 17\%$ of the total baryon density \citep{Fukugita1998}. Hence, the amount of compact objects in star is small, i.e. $\sim 0.025$.  Thus, these constraints put a stringent upper limit on the fraction of DM as compact objects of 50$\%$. In addition, we find $\Omega_{\rm M} = 0.304 \pm 0.0067$, consistent with the fiducial case in the literature which assumes homogeneity \citep[e.g.][]{Brout2022}. 
For the fiducial case, we use a prior on $\eta$ of U[0, 1], however, as we find that the posterior distribution of $\eta$ peaks close to 1, we test what constraints we get when using a wider prior of U[0, 5]. We find that the limit on $f_p$ in this case is 0.32 at the 95\% C.L. Therefore, the limit on $f_p$ from the asymmetric posterior distribution of $\eta$ is a conservative upper limit on the compact object DM fraction.

\vspace{-0.4cm}
\subsection{Alternate Cosmological Models}
The constraints on $f_p$ presented here assume the standard flat $\Lambda$CDM model, i.e. with only $\Omega_{\rm M}$ as a free parameter. We test the impact of using extended cosmological models with additional free parameters on the inferred $f_p$ values. 
\vspace{-0.3cm}
\subsubsection{$w$CDM}
The $w$CDM model is the simplest phenomenological extension of the $\Lambda$CDM cosmology, with the dark energy equation of state (EoS) being a free parameter in the model, assumed to be a constant over redshift. In this model we find $\eta > 0.62$ at the 95\% C.L., implying that $f_p < 0.49$. We note that the value of $w$ is weakly correlated with $\eta$ which worsens the constraints on the $f_p$ value compared to the case with $w=-1$. We constrain $w = -0.954 \pm 0.04$, a $\sim 5\%$ constraint on the dark energy EoS. Compared to the case with a fixed $\eta = 1$, these constraints are slightly weaker, however, they are consistent with the homogeneous case, as studied in \citet{Brout2022}. This shows that the impact of inhomogeneities on the inferred value of $w$ is small, consistent with previous works \citep{Dhawan2018c}. These constraints are consistent with the fiducial $w$CDM values presented in \citet{Brout2022} We also note that in this model, $\eta$ is positively correlated with $H_0$, suggesting that $\eta$ is closer to 1 for faster expansion.  We note that the constraints on $\Omega_{\rm M}$ in this model are slightly lower than for the fiducial $\Lambda$CDM case, however, they are within the $\sim 1 \sigma$ error region. This is consistent with the $ w > -1$ estimate of the equation of state for this model. 
\vspace{-0.3cm}
\subsubsection{o$\Lambda$CDM}
In the o$\Lambda$CDM, we do not assume $\Omega_{\rm K}=0$ and instead constrain it from the combination of the BAO, CMB and SN~Ia data. We find $\Omega_{\rm K} = -0.0022 \pm 0.0021$ and an upper bound on the fraction of compact objects of $f_p =0.47$. 
Constraints from all the cosmological models are summarised in Table~\ref{tab:constraints}. 
\vspace{-0.3cm}
\subsection{Impact of $z$ distribution}
We test the impact of the high-$z$ SNe~Ia in constraining $f_p$ by analysing subsamples of the Pantheon+ catalog with a $z_{\rm max}$ selection from 0.5 to 1. We find that there are no strong constraints on $f_p$ at the 95 \% C.L. when the upper limit of the $z$ distribution is 0.5. With $z_{\rm max} \geq 0.8$,  $f_p$ at 95\% C.L. is constrained to 0.43. This demonstrates the importance of the highest-$z$ SNe~Ia in placing limits on $f_p$. 

We note, however, that the most stringent constraints are not obtained from the entire sample but rather from the subsample truncated at $z \leq 1$. We surmise that this could be due to additional systematic uncertainties in the small sample of $z > 1$ SNe~Ia which are causing the constraints to be weakened compared to the complete sample or a statistical chance event due to the small sample size in the highest-$z$. However, for this study we consider the conservative case of the entire sample as default.
This will be an interesting investigation with future high-$z$ samples, which are uniformly constructed from a single SN survey e.g. from the Vera C. Rubin Observatory \citep{SRD_LSST} or Roman Space Telescope  \citep{Hounsell2018}. 

\begin{figure}
    \centering
    \includegraphics[width=.5\textwidth]{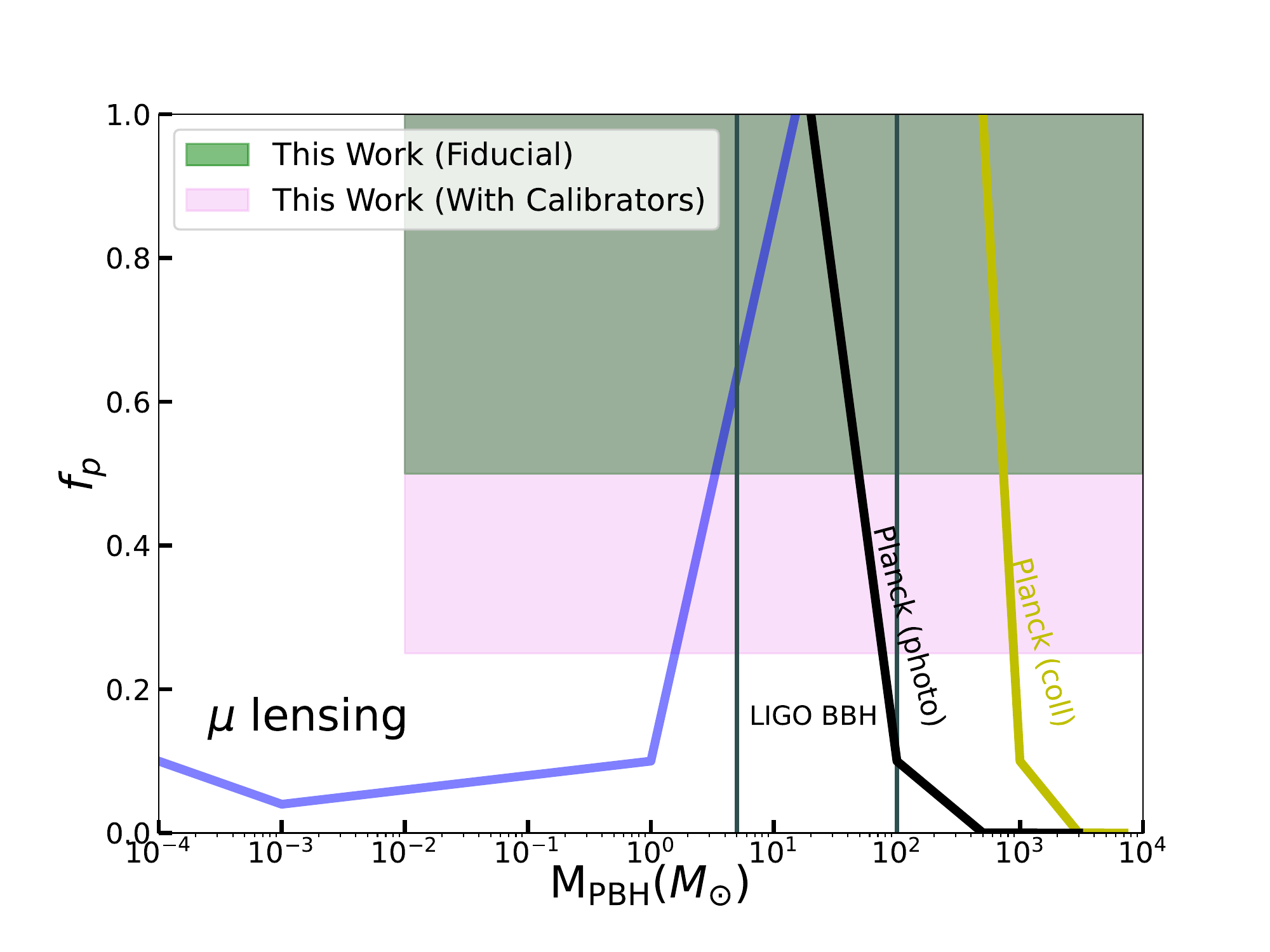}
    \caption{Constraints on primordial black holes as dark matter for different PBH mass range. The microlensing constraints (blue) are predominantly strong only for very light PBHs whereas the CMB (black and yellow solid lines) mainly constrain higher mass PBHs than the SN~Ia constraints in this work (green and pink). The LIGO BBH mass range is shown for comparison. We find that when including the calibrators, the constraints on $f_p$ from SNe~Ia only allow for a maximum of 25\% of DM to be composed of PBHs.   }
    \label{fig:exclusion_fp}
\end{figure}
\vspace{-0.4cm}
\subsection{Constraints with Calibrator SNe~Ia}

In our fiducial analyses we do not include calibrator SNe~Ia, i.e. SNe~Ia for which the absolute magnitude have been constrained using local distance ladder measurements, thus avoiding the combination of datasets that are in tension with each other. Here, we test the impact of adding constraints on the cosmological parameters from the calibrator SNe~Ia as well. We follow the procedure outlined in \citet{Dhawan2020,Brout2022} where the calibrator SNe~Ia constrain $M_B$ (and hence, $H_0$) and the Hubble flow SNe~Ia with $z \geq 0.023$ constrain the shape of the high-$z$ magnitude-redshift relation. We use the complete covariance matrix as provided by \citet{Brout2022} and find an upper limit $\eta > 0.81$ 95\% C.L. corresponding to an upper limit on $f_p$ of 0.25 at the 95\% C.L. The inferred Hubble constant is $H_0 = 69.18 \pm 0.50$ km\,s$^{-1}$\,Mpc$^{-1}$. We note that these are the tightest constraints on $f_p$, and very significantly rule out PBHs as the dominant DM component.
For comparison, we fit (to combined probes)  the cosmological parameters with $\eta = 1$ , i.e., the case assuming homogeneity, and find $H_0 = 69.29 \pm 0.49$ km\,s$^{-1}$\,Mpc$^{-1}$, consistent with the fiducial case including inhomogeneities. We conclude that the inclusion of possible small scale inhomogeneities have a negligible effect on the inferred value of $H_0$.
\vspace{-0.6cm}
\section{Discussion}
\label{sec:discuss}
SNe~Ia have previously been used to constrain $\eta$ within the DR distance framework, to test for the impact on dark energy properties \citep{Dhawan2018c}. While the assumption of homogeneity doesn't bias dark energy properties, the previous datasets did not provide stringent constraints on the amount of DM in compact objects. The method of using the DR distance to compute the fraction of DM in compact objects is a simple approach with very few model assumptions. The constraints on $\eta$, the homogeneity parameter, are independent of any modelling assumptions on the matter distribution. For inferring, $f_p$, we use the scaling relation from \citet{Mortsell2002}, which, although using numerical simulations, finds the relation to be insensitive to the choice of inhomogeneity and cosmological parameters. This work presents the first strong constraints on compact object DM using the DR distance method. 

Previously, the SN~Ia Hubble diagram had been used to constrain $f_p$ with the magnification probability density function (PDF). This is another powerful route to constrain $f_p$ from the SN~Ia Hubble diagram.  These constraints are typically stronger on $f_p$ \citep[e.g., see comparison with previous DR distance constraints][]{Dhawan2018c}, although requiring more model assumptions.  The constraints are derived due to two effects on the magnification distribution. Firstly, the fact that most light ray will pass far away from the compact objects, hence, will be demagnified compared to the mean, secondly that a small fraction will pass by close to the compact objects leading to a long tail towards high magnification, i.e. overluminous outliers. The method is susceptible to be conflated with the presence of intrinsically overluminous SNe~Ia, e.g. the SN~1991T-like subclass \citep{Phillips2022} and hence, weaken the constraints on $f_p$ \citep[e.g., see][]{ZS2018}.  The simplicity of the DR distance approach  allows for robust marginalisation over cosmological parameters under several different model assumptions, and is expected to be fairly insensitive to uncertainties in the high magnification tail of the SN~Ia magnitude-redshift relation. Therefore, it is a useful complement to the magnification PDF analyses.

\vspace{-0.5cm}
\section{Conclusion}

In our fiducial case of the flat $\Lambda$CDM, we find an upper limit of $f_p < 0.50$ at the 95\% C.L. When we assume a wide prior on $\eta$  between 0 and 5 such that the prior does not truncate the posterior if the data prefers $\eta \sim 1$, we find that the upper limit on $f_p$ is 0.32 at 95\% C.L., the most stringent limit on $f_p$ till date.
We also infer $f_p$ in $w$CDM and o$\Lambda$CDM models where the dark energy EoS and curvature are free parameters. In both cases we find very similar limits on $f_p$ compared to the fiducial case, suggesting that the constraints are robust to assumptions of the cosmological model. We note that since this constraint on $f_p$ also includes the compact object budget in stars and stellar remnants. While this is a minor contribution to the total compact object budget \citep[e.g., see][]{Fukugita1998},  these constraints are, therefore, conservative upper limits .

Comparing the constraints on $H_0$ from the cases with $\eta$ as a free parameter and $\eta$ fixed to 1, i.e. the homogeneous case, we find no significant differences and hence conclude that the inhomogeneities cannot explain a significant part of the observed tension.
We find a marginal increase in the uncertainty on $w$ compared to the case assuming homogeneity. Moreover, the value of $w$ is not biased compared to the homogeneous case. 
 


We compare the constraints from our work with other probes of the abundance of PBHs in various mass ranges in Figure~\ref{fig:exclusion_fp}. While microlensing constraints are stringent in the regime of solar mass PBHs and below \citep{Tisserand2007}, within the range of masses where LIGO has discovered binary BHs \citep{Ligo_2022}, the tightest constraints are from the SN~Ia Hubble diagram. For larger masses, there are strong constraints from CMB observations \citep{Bernal2017}. 
 We note, however, that even if PBHs do not constitute a dominant fraction of DM, sufficiently large PBHs might generate cosmic structure and possibly supermassive BHs in galactic nuclei \citep[e.g., see][]{Carr_2022}, hence, its important to have precise constraints on their cosmological density.

We find that this approach presents a robust and stringent constraints on the fraction of DM in PBHs, strongly suggesting that not all DM is composed of compact objects with M$\gtrsim 0.01 {\rm M_{\odot}}$, indicating the need for a new elementary particle as a DM candidate. 
\vspace{-0.7cm}
\section*{Acknowledgements}
We thank Phil Bull for stimulating discussions. SD acknowledges support from the Marie Curie Individual Fellowship under grant ID 890695 and a Junior Research Fellowship at Lucy Cavendish College. EM acknowledges support from the Swedish Research Council under contract 2020-03384.
\vspace{-0.8cm}
\section*{Data Availability}
The data used in this paper are publically available and the analysis was performed using standard inference softwares. Code and datasets will be made available upon reasonable request.


\vspace{-0.5cm}
\bibliographystyle{mnras}
\bibliography{main} 








\bsp	
\label{lastpage}
\end{document}